\documentclass[prl,aps,twocolumn,10pt, a4paper,longbibliography]{revtex4-2}

\usepackage[paperwidth=210mm,paperheight=297mm,centering,top=2cm,bottom=2.8cm,right=1.8cm,left=1.8cm]{geometry}

\pdfoutput=1

\date{\today}

\usepackage{mathrsfs} 
\usepackage{xspace}

\newcommand{\um}{\upmu{\rm m}}

\newcommand{\kB}{k_{\textrm{B}}}
\newcommand{\kD}{k_{\textrm{D}}}

\newcommand{\UD}{U_{\textrm{D}}}
\newcommand{\tD}{t_{\textrm{D}}}

\newcommand{\kxi}{k_\xi}
\newcommand{\mueq}{\zeta}

\usepackage{float}
\usepackage{soul}
\usepackage{fourier}

\usepackage[dvipsnames]{xcolor}
\usepackage{graphicx}
\usepackage{amsmath,amssymb}

\usepackage{times}
\usepackage{upgreek}
\usepackage{psfrag} 
\usepackage{latexsym} 
\usepackage{amstext}
\usepackage{amsxtra} 

\usepackage{}

\usepackage{lineno}

\usepackage{textcomp}
\usepackage{amsfonts}
\usepackage{graphicx}
\usepackage{bm}
\usepackage{color}
\usepackage{braket}
\usepackage{siunitx}
\usepackage[svgnames]{xcolor}
\usepackage[normalem]{ulem}

\definecolor{myColor}{rgb}{0.02,0.12,0.3}
\definecolor{myciteColor}{rgb}{0.39,0.7,0.89}
\definecolor{myBlue}{rgb}{0.1,0.1,0.8}
\usepackage[colorlinks=true,citecolor=myColor,linkcolor=myColor,urlcolor=myColor]{hyperref}

\def\be{\begin{equation}}
\def\ee{\end{equation}}

\makeatletter
\def\@fnsymbol#1{\ensuremath{\ifcase#1\or *\or \dagger\or \ddagger\or
   \mathsection\or \mathparagraph\or \|\or **\or \dagger\dagger
   \or \ddagger\ddagger \else\@ctrerr\fi}}
\makeatother

\usepackage{titlesec}

\begin{document} 
\title{A Nonequilibrium Equation of State for a Turbulent 2D Bose Gas}

\author{Yi Jiang$^*$, Nikolai Maslov$^*$, Andrey Karailiev, Christoph Eigen, Martin Gazo$^{\dag}$, and Zoran Hadzibabic}
\affiliation{Cavendish Laboratory, University of Cambridge, J. J. Thomson Avenue, Cambridge CB3 0US, United Kingdom}

\begin{abstract}

Nonequilibrium equations of state can provide an effective thermodynamic-like description of  far-from-equilibrium systems.
We experimentally construct such an equation for a direct energy cascade in a turbulent two-dimensional Bose gas. Our homogeneous gas is continuously driven on a large length scale and, with matching dissipation on a small length scale, exhibits a nonthermal but stationary power-law momentum distribution. Our equation of state links the cascade amplitude with the underlying scale-invariant energy flux, and can, for different drive strengths, gas densities, and interaction strengths, be recast into a universal power-law form using scalings consistent with the Gross--Pitaevskii model.

\end{abstract}

\maketitle

The framework of thermodynamics provides an effective description of many-particle systems at or near equilibrium, encapsulating complex microscopic behavior with a small number of macroscopic state variables that are related by a system-dependent equation of state (EOS)~\cite{Chaikin:1995,Borsanyi:2014,Lattimer:2016,Pitaevskii:2016,Gaffney:2018}. There is a lot of interest, and also progress, in transposing these ideas to far-from-equilibrium systems that may not attain even local thermodynamic equilibrium, including active matter~\cite{Ginot:2015,Takatori:2015,Solon:2015}, glasses~\cite{Cugliandolo:1997}, and plasmas~\cite{Cassak:2023}.

\begin{figure}[b]
\centering
\includegraphics[width=\columnwidth]{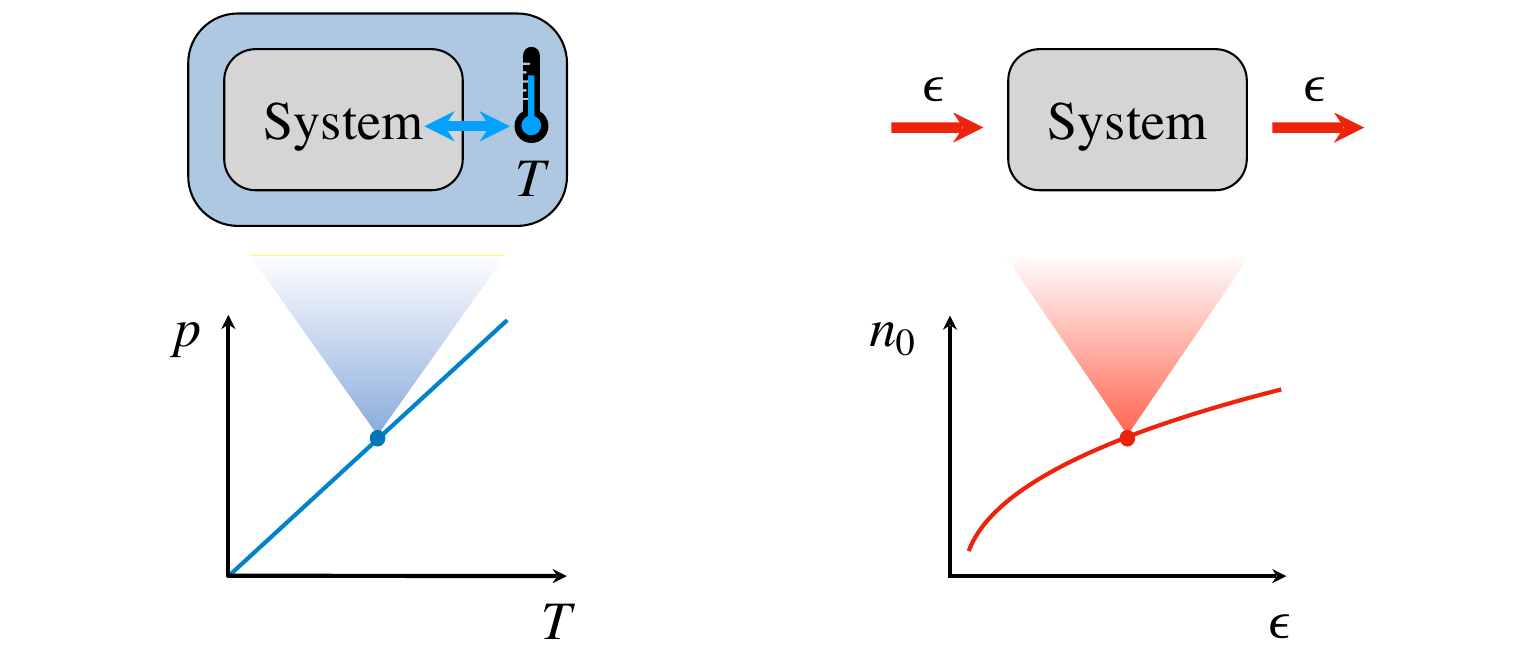}
\caption{
The equation of state (EOS) for an equilibrium system (left) and a flux-carrying turbulent system in a steady state, with matching injection and dissipation rate $\epsilon$ (right).
For an equilibrium system, coupled to a reservoir, state variables such as pressure, $p$, are functions of the reservoir temperature $T$.
For a flux-carrying turbulent system, the amplitude, $n_0$, of the emergent power-law spectrum is a function of the underlying flux $\epsilon$.
}
\label{fig1}
\end{figure}

Driven turbulent cascades exhibit stationary far-from-equilibrium states, which are natural candidates for a thermodynamic-like characterization. They feature scale-invariant transport of conserved quantities, such as energy~\cite{Richardson:1922,Kolmogorov:1941,Obukhov:1941} or enstrophy~\cite{Kraichnan:1967b}, between an injection scale, $k_{\rm I}$ (where $k$ is a wavenumber), and a dissipation scale, $\kD$. In the inertial range, between $k_{\rm I}$ and $\kD$, spectra of system-dependent quantities such as energy or wave action generically exhibit power-laws with universal exponents~\cite{Kolmogorov:1941,Obukhov:1941,Kraichnan:1967b,Nazarenko:2011,Schekochihin:2022,Barenghi:2023b,Zakharov:2025}.
The state of the system (in this $k$-range) is then specified by the amplitude of the (statistically averaged) power-law spectrum. The relation of this amplitude to the underlying scale-invariant flux~\cite{Kolmogorov:1941,Obukhov:1941,Kraichnan:1967b,Sreenivasan:1995,Nazarenko:2011,Schekochihin:2022,Barenghi:2023b,Dogra:2023,Zakharov:2025} is analogous to an EOS in equilibrium thermodynamics~\cite{Dogra:2023}, but with intrinsically nonequilibrium state variables; see Fig.~\ref{fig1}.

In this Letter, we experimentally establish a nonequilibrium EOS for a direct energy cascade (from large to small length scales, {\it i.e.}, with $k_{\rm I}<\kD$) in a homogeneous two-dimensional (2D) Bose gas~\cite{Galka:2022}, which in a turbulent steady state exhibits a power-law momentum distribution (wave-action spectrum) $n_k$ with a universal exponent $\gamma$ and an amplitude $n_0$.
Using scalings consistent with the Gross--Pitaevskii model, we show that our measurements for different drive strengths, gas densities, and strengths of the interparticle interactions, can be recast into a universal power-law relation, $n_0 \propto \epsilon^b$, where $\epsilon$ is the cascade energy flux and $b=0.47(5)$.
Our measured $\epsilon$ satisfies the zeroth law of turbulence~\cite{Frisch:1995,Vassilicos:2015,Navon:2019} and, by quenching the drive strength, we verify that the steady-state $n_0$ satisfies the same EOS independently of the system's history.

\begin{figure*}[t]
\centering
\includegraphics[width=\textwidth]{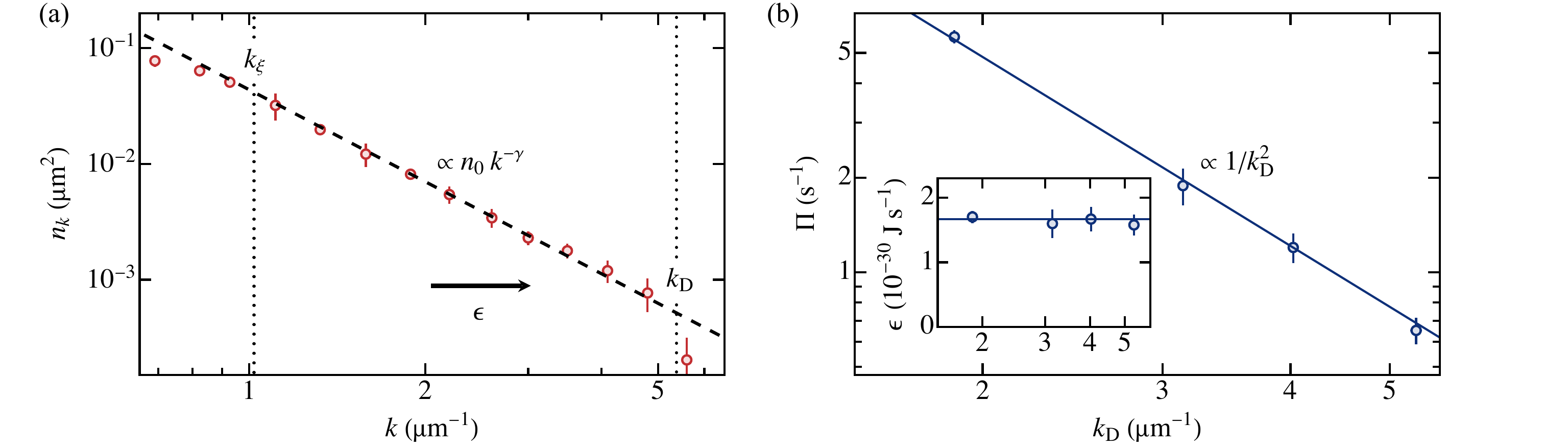}
\caption{Turbulent steady state and the underlying energy flux. (a) An example of a steady-state momentum distribution $n_k(k)$ for our turbulent 2D gas, held in a square box trap of side length $L=\SI{30}\,{\um}$ and driven on a large length scale ($\sim L$) by a time-periodic force. Between the inverse healing length $\kxi$ and the dissipation scale $\kD$, set by our trap depth, $n_k$ exhibits a power-law spectrum, $\propto n_0\,k^{-\gamma}$, with $\gamma\approx 2.6$ (dashed line). Here, the gas has density $n = 25\, \si{\um^{-2}}$ and dimensionless interaction strength $\tilde{g} = 0.04$, and the force amplitude is $F_0 = 1.6 \,\zeta/L$, where $\zeta = \hbar^2\tilde gn/m$. (b) The corresponding energy flux $\epsilon$. The steady-state particle flux $\Pi$ is measured through atom loss for various $\kD$. In agreement with the zeroth law of turbulence, we observe $\Pi \propto 1/\kD^2$ (solid line), and thus deduce the $\kD$-independent $\epsilon = \Pi\,\hbar^2\kD^2/(2m)$ (inset). Note that $n_k$, $\Pi$, and $\epsilon$ are normalized per particle, and the spectrum in (a) was measured with the largest $\kD$ value in (b).}

\label{fig2}
\end{figure*}

We start with a quasipure homogeneous 2D condensate of $^{39}$K atoms in the $|F=1,m_F=-1\rangle$ hyperfine state, confined to the $x$-$y$ plane by a harmonic trap with frequency $\omega_z= 2\pi \times \qty{1.2}{\kilo \hertz}$, and within the plane by a square optical box trap of side length $L=\qty{30}{\micro \metre}$ and depth $\UD=\kB\times \qty{175}{\nano \kelvin}$~\cite{Christodoulou:2021}. For a 2D gas, the interaction strength is given by the dimensionless $\tilde g = a\sqrt{8 \pi m \omega_z / \hbar}$~\cite{Petrov:2000a,Hadzibabic:2011}, where $m$ is the atom mass and $a$ is the \textit{s}-wave scattering length, which we tune using a Feshbach resonance at $33.6\,$G~\cite{Chapurin:2019}. For our variable atom number, $N= (20-36)\times 10^3$, corresponding to density $n=N/L^2 = (22-40)\, \si{\um^{-2}} $, and $\tilde{g} = (0.02-0.08)$, the characteristic interaction energy per particle is $\zeta=\hbar^2\tilde gn/m$.

We continuously inject energy into the gas by resonantly driving a lowest-energy phonon mode, with wavenumber $k_L=\pi/L\approx \SI{0.1}{\per\um}$ and angular frequency $\omega_{\rm B} = \sqrt{\zeta/m} \,\,k_L$, using a spatially uniform time-periodic force along one of the box axes~\cite{Navon:2016,Christodoulou:2021, Galka:2022}; for our different measurements, we vary the force amplitude $F_0$ in the range $(0.7-12)\,\zeta/L$.
This injection is anisotropic, but the steady-state momentum distribution $n_k$ shows an emergent statistical isotropy for $k \gtrsim k_\xi$, where $k_\xi = \sqrt{\tilde g n} =(0.7-1.8)\,{\um^{-1}}$ is the inverse healing length~\cite{Galka:2022}; from a theoretical point of view~\cite{Zhu:2023,Martirosyan:2026,Zhu:2025}, $k_\xi$ is the effective injection scale, $k_{\rm I}$, for a statistically isotropic cascade.
The continuous energy dissipation occurs in the form of a small particle-loss rate due to atoms escaping the trap at $\kD = \sqrt{2 m \UD} / \hbar$~\cite{Navon:2019,lossfootnote2dEOS}; for our full trap depth $\kD \approx \SI{5.3}{\per\um}$, but note that for some measurements we also reduce $\kD$.

In Fig.~\ref{fig2}(a), for one set of system parameters, we show the azimuthally averaged $n_k$ for a steady-state turbulent cloud; we measure $n_k$ using matter-wave focusing~\cite{Tung:2010,Gazo:2025} and normalize it such that $\int n_k \, 2\pi k\, {\rm d}k=1$.
For $k_{\xi}\lesssim k\lesssim \kD$, we observe a power-law $n_k \propto n_0 \,k^{-\gamma}$, with $\gamma=2.6$ (dashed line). In the theory of weak-wave turbulence (WWT), the Kolmogorov--Zakharov (KZ) analytical solution for a direct energy cascade in a Bose gas predicts (for $k \gtrsim \kxi$) a power-law $n_k$ with $\gamma = d$, where $d$ is the system dimensionality~\cite{Nazarenko:2011,Zakharov:2025}. 
However, this solution is not valid for $d=2$ because $n_k \propto k^{-2}$ coincides with the equilibrium (Rayleigh--Jeans) distribution and carries no flux.
In the various theories that address this problem~\cite{Malkin:1996,Nazarenko:2011,Falkovich:2015,Skipp:2020}, $n_k$ is, over finite $k$-range, close to a power law with $\gamma>2$ (see also~\cite{Zhu:2023}).

In Fig.~\ref{fig2}(b), we show our measurement of the corresponding energy flux $\epsilon$, deduced from the scale-invariant particle flux $\Pi$~\cite{Navon:2019}; note that both fluxes are normalized per particle.
The main panel shows measurements of $\Pi$, which is directly seen in the rate at which atoms leave the trap at $\kD$; ${\Pi} = - \dot{N}/N$. We measure $\Pi$ for various $\kD$ (by reducing the trap depth) and, in agreement with the zeroth law of turbulence~\cite{Frisch:1995,Vassilicos:2015}, observe $\Pi \propto 1/\kD^2$ (solid line), which corresponds to a $\kD$-independent $\epsilon = \Pi \,\hbar^2 \kD^2/(2m)$ (see inset).

\begin{figure*}[t]
\centering
\includegraphics[width=\textwidth]{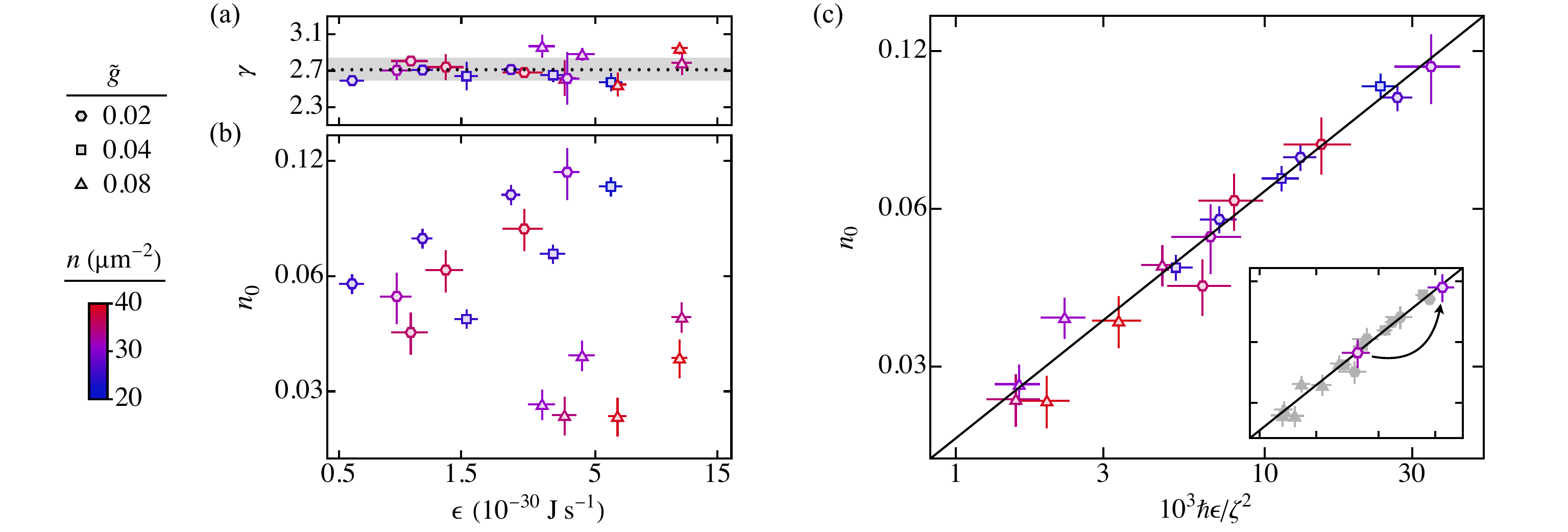}
\caption{Universal EOS. Here we generalize our measurements of the steady-state $n_k$ to various $\tilde{g}$, $n$, and $\epsilon$.
(a) The cascade exponent~$\gamma$. Our data show no systematic variation and are consistent with $\gamma = 2.7(1)$ (dotted line and shading).
(b) The cascade amplitude $n_0$ (extracted with fixed $\gamma=2.7$) plotted versus $\epsilon$.
(c) Plotting $n_0$ versus the dimensionless energy flux $\hbar\epsilon/\mueq^2$ collapses all our data onto a universal EOS. The solid line shows a power-law fit, with exponent $b = 0.47(5)$. Inset: we highlight the two data points corresponding to the steady states established before and after a change in the driving-force amplitude (see text).
}
\label{fig3}
\end{figure*}

In Fig.~\ref{fig3}, we summarize our measurements for various $\tilde{g}$, $n$, and $\epsilon$, and construct a universal EOS.
Note that here we always measure $n_k$ for our largest $\kD$ and that, for fixed $\tilde{g}$ and $n$, we vary $\epsilon$ using $F_0$ as the experimental control parameter. The emergent $\epsilon$ is monotonic in $F_0$, but their relationship is in general complicated, and to construct the universal EOS, one needs to relate $n_0$ to the measured $\epsilon$ rather than the externally imposed $F_0$.

In Fig.~\ref{fig3}(a) we show that we always observe $\gamma$ consistent with $2.7(1)$~(see also~\cite{Galka:2022}), without any systematic dependence on $\tilde{g}$, $n$, or $\epsilon$.
From here on, we define a dimensionless $n_0 = \kxi^{2-\gamma} \langle k^{\gamma} n_k\rangle$~\cite{Dogra:2023} with fixed $\gamma=2.7$ and always take the average $\langle k^{\gamma} n_k\rangle$ for $k$ in the range $(2-4) \,\si{\per\um}$. 

In Fig.~\ref{fig3}(b), we show all our measurements of $n_0(\epsilon)$ for different $\tilde{g}$ and $n$, and in Fig.~\ref{fig3}(c) we show that they all collapse onto a single universal curve when we plot $n_0$ versus the dimensionless energy flux $\hbar\epsilon/\mueq^2$ (using a normalization of $\epsilon$ that is natural within the Gross--Pitaevskii model~\cite{Dogra:2023}). This universal EOS is captured well by a power-law fit, $n_0 \propto (\hbar\epsilon/\mueq^2)^b$ (solid line), with $b=0.47(5)$.

Finally, for $n_0$ and $\epsilon$ to be analogous to equilibrium state variables, their relationship should be independent of the history of the system. To verify this, we performed an experiment where we prepared a turbulent steady state and then changed $F_0$ (from $2\,\mueq/L$ to $6\, \mueq/L$), confirming that the steady-state $n_0$ before and after this change satisfy the same EOS; these two measurements are included in Fig.~\ref{fig3}, and are highlighted in the inset of Fig.~\ref{fig3}(c). 

Theoretically, the problem of the functional form of the EOS in 2D is still quite open. 
Some of the theories with flux-carrying $n_k$ predict $b=1/3~\cite{Nazarenko:2011,Skipp:2020}$, while others predict $b=1/2$~\cite{Malkin:1996,Falkovich:2015} (see also~\cite{Zhu:2025}), which is physically associated with a significance of (effective) 3-wave interactions in a Bose gas. The latter prediction agrees with our experiments, but this could be fortuitous. For example, beyond-WWT effects, such as vortex excitations, could also play a role, as seen in numerical studies of 3D Bose gases~\cite{Fischer:2025, Martirosyan:2026}.

Our results provide benchmarks for theories of quantum-gas turbulence~\cite{Nazarenko:2011,Barenghi:2023b,Zakharov:2025} and contribute to establishing a general framework of nonequilibrium thermodynamics~\cite{Cugliandolo:1997,Ginot:2015,Takatori:2015,Solon:2015,Dogra:2023,Cassak:2023}.
In the future, it would be interesting to extend studies of far-from-equilibrium EOS to inverse cascades and to systems with fermionic statistics or long-range interactions. Moreover, in analogy with the fluctuation-dissipation relations in equilibrium thermodynamics, it would be interesting to explore connections between the underlying flux and the fluctuations in $n_k$ around the steady state.

This work was supported by ERC [UniFlat], EPSRC [Grant No.~EP/Y01510X/1], and STFC [Grants No.~ST/T006056/1 and No.~ST/Y004469/1].
Z.~H. acknowledges support from the Royal Society Wolfson Fellowship.

%


\clearpage
\setcounter{figure}{0} 
\setcounter{equation}{0}
\renewcommand\theequation{S\arabic{equation}} 
\renewcommand\thefigure{\arabic{figure}} 
\renewcommand{\figurename}[1]{Extended Data Fig.~}

\titlespacing\subsection{0pt}{11pt}{11pt}

\end{document}